\documentclass{article}

\usepackage[left=1in,top=1in,right=1in,bottom=1in,nohead]{geometry}
\usepackage{float, multirow, amsmath, amssymb, graphicx, epstopdf, qtree}
\usepackage{cite}
\usepackage[english]{babel}
\usepackage{url}
\usepackage{color,soul}
\soulregister\cite{7}
\renewcommand{\hl}[1]{#1}		

\title{Lexical representation explains cortical entrainment during speech comprehension}

\author{Stefan L. Frank\\Centre for Language Studies, Radboud University\\ \\Jinbiao Yang\\Institute of Brain and Cognitive Science, NYU Shanghai} 
\date{}

\begin{document}
\maketitle

\begin{abstract}
Results from a recent neuroimaging study on spoken sentence comprehension have been interpreted as evidence for cortical entrainment to hierarchical syntactic structure. We present a simple computational model that predicts the power spectra from this study, even though the model's linguistic knowledge is restricted to the lexical level, and word-level representations are not combined into higher-level units (phrases or sentences). Hence, the cortical entrainment results can also be explained from the lexical properties of the stimuli, without recourse to hierarchical syntax. 
\end{abstract}

\section{Introduction}
There is considerable debate on the precise role of hierarchical syntactic structure during sentence comprehension, with some arguing that full hierarchical analysis is part and parcel of the comprehension process (e.g., \cite{Berwick2013,Hale2011}, among many others) and others claiming that more shallow \cite{Sanford2002} or even non-hierarchical \cite{Frank2012,Jackendoff2017} processing is common.

Ding, Melloni, Zhang, Tian, and Poeppel \cite{Ding2016} recently 
presented evidence that cortical entrainment during speech perception reflects the neural tracking of hierarchical structure of simple sentences, which would support the view that hierarchical processing is inevitable. 
They had participants listen to sequences of linguistic material consisting of English monosyllabic words or Chinese syllables, presented at a fixed rate of one syllable every 250~ms (or at a slightly slower rate for English). Depending on the experimental condition, each four-unit subsequence  contained linguistic units at one or two hierarchically higher levels, or lacked meaningful structure beyond the syllable. For example, part of the stimulus in the English four-word 
sentence condition could be ``... dry fur rubs skin fat rat sensed fear ...'' where a group of four consecutive words forms a sentence with
the following hierarchical structure:

(1) \Tree [.S [.NP [.Adj dry ] [.N fur ] ] [.VP [.V rubs ] [.N skin ] ] ] 

\vspace{\baselineskip}
Here, the labels Adj, N, and V stand for the syntactic categories adjective, noun, and verb, respectively; NP and VP represent noun phrase and verb phrase; and S denotes the complete sentence. 

Magnetoencephalography (MEG) signals were recorded from participants listening to such a word sequence. Power spectra computed from these signals displayed peaks at precisely the presentation rates of words, phrases, and sentences. Ding et al. interpreted this finding as evidence for neural tracking of language at these three, hierarchically related levels. A similar result was obtained from Chinese sentences consisting of a two-syllable noun followed by a two-syllable verb or verb phrase. In this condition, the power spectrum showed peaks at the occurrence frequency of the syllable, the word/phrase, and the sentence; that is, at 4~Hz, 2~Hz, and 1~Hz, respectively. When the sentence-level structure was removed by presenting only the Chinese verbs and verb phrases, only the 4~Hz and 2~Hz peak remained; and in a shuffled-syllable condition without any \hl{consistent} lexical or sentential structure, the power spectrum was reduced to a 4~Hz peak only. Finally, when sentences consisted of a monosyllablic verb followed by a three-syllable noun or noun phrase, there were significant peaks in the power spectrum at 4~Hz and 1~Hz but not at 2~Hz, as there was no longer any linguistic unit occurring at a 2~Hz rate. 

To summarize the Ding et al. results: Across conditions, spectral power peaks occur exactly at the presentation frequencies of the linguistic units at the various hierarchical levels present in the stimulus. However, this does not yet imply a causal relation between cortical entrainment and the processing of hierarchical syntactic structure. As we will demonstrate by means of a simple computational model that does not incorporate syntactic structures or any other linguistic knowledge beyond the word level, the same entrainment results can follow from lexical representation alone. 

Our model represents the stimuli from the Ding et al. experiments as sequences of high-dimensional numerical vectors. These are assigned by a distributional semantics model (trained on large amounts of English or Chinese text) such that two words that tend to occur in similar local contexts receive similar vectors. Consequently, similarities between vectors reflect similarities between the semantic/syntactic properties of the represented words. For example, the inanimate nouns ``fur'' and ``skin'' will have similar representations, which will differ somewhat different from the vector for the animate noun ``rat'', which in turn will be quite distinct from the vector for the action verb ``rubs''. 
Psycholinguists have argued that information about linguistic distributions partly underlies knowledge of word meaning \cite{Andrews2009}. Distributional semantics is also increasingly influential in semantic theory \cite{Baroni2014,Coecke2010} and widely applied in  computational linguistics where it yields state-of-the-art results in applications for natural language processing \cite{Bellegarda2016}. Vector representations for words account for experimental findings from psycholinguistics \cite{Mandera2017,Rotaru2017} and are not unrelated to cortical representations: They have been shown to allow for the decoding of neural activity during single word comprehension \cite{Mitchell2008} or narrative reading \cite{Wehbe2014PLoS,Wehbe2014EMNLP}, and distances between vectors are predictive of neural activation during written and spoken language comprehension \cite{Frank2017,Li2016}. We use these vectors differently here: Rather than comparing vector (distances) to neural activation, 
we compute power spectra directly over sequences of vectors, as Ding et al. do for recorded MEG signals.
Vector sequences that represent the Ding et al. stimuli in the different conditions result in \hl{power spectra that are very similar to those from} human participants. Hence, the cortical entrainment results need not be indicative of the detection or construction of hierarchical structures.

\section{Methods}

\subsection{Materials}
All materials were taken directly from the Ding et al. experiments. For English, these were 60 four-word sentences with structure as in (1) except that in seven sentences, the first word was not an adjective but a numeral or possessive pronoun. All English words were monosyllabic.
For Chinese,  the written items provided by  Ding et al. were converted into pinyin (a phonological representation of Mandarin Chinese) by Pypinyin (version 0.12.1; we replaced the package's word-pinyin dictionary with a larger one from ZDIC) after which the result was manually checked and corrected when needed.
Converting the written characters to pinyin is essential because the original experiments used auditory stimuli presentation.
The same pinyin syllable can correspond to many different written characters, making word co-occurrence patterns differ strongly between spoken and written Chinese. 
Consequently, a distributional semantics model can only adequately capture the lexical information in spoken stimuli if it is applied to the pinyin form. 

There were two sets of 50 four-syllable Chinese sentences in the  Ding et al. experiments. Sentences in the first set consisted of a two-syllable noun and a two-syllable verb or verb phrase, for example ``l\v{a}oni\'{u} g\={e}ngd\`{i}''(``Old cattle ploughs the field''). 
The second set of Chinese sentences consisted of a monosyllablic verb followed by a three-syllable noun or noun phrase, for instance, ``zh\={e}ng  gu\`{a}nt\={a}ngb\={a}o''(``Braising the soup dumplings''). 

Following  Ding et al., syllable strings containing only two-syllable verbs or verb phrases were constructed by taking the verb (phrase) parts from the [N V(P)] sentences. Furthermore, four-syllable sequences without any consistent word, phrase, or sentence structure were obtained by randomly reassigning syllables to the [N V(P)] sentences while retaining each syllable's position in the sentence. 

To summarize, there are five experimental conditions using the same stimuli as  Ding et al.: one in English ([NP VP] sentences) and four in Chinese ([N V(P)] sentences, [V N(P)] sentences, verb (phrases) only, and shuffled syllable sequences). Depending on condition, each stimuli sequence is composed of either four-syllable sentences, two-syllable verbs or verb phrases, or individual syllables without further linguistic structure. 

\subsection{Representing lexical knowledge}

\subsubsection{Distributional semantics}
Vector representations of English and Chinese words were generated by the Skipgram distributional semantics model \cite{Mikolov2013}, which is a feedforward neural network with $N$ hidden units and input/output units that each represent a word type. The network is exposed to a large amount of English or Chinese text and, for each word token, learns simultaneously to predict the five following words and to retrodict the five preceding words. Words that are paradigmatically related to one another tend to occur in similar contexts, resulting in similar weight updates in the network. Consequently, connection weights come to represent words such that they capture paradigmatic relations among the represented words. More precisely, after training, a word is represented by the $N$-dimensional weight vector of connections emanating from that word's input unit, and words with similar syntactic or semantic properties will have similar vectors. Words that occurred less than five times in the training corpus were excluded to reduce processing time and memory requirements, and because the distributional information for infrequent words is less reliable.

\hl{For each language,} we obtained twelve different sets of vectors (i.e., simulated twelve participants) by running the Skipgram model twelve times with hidden-layer size $N$ randomly drawn from a normal distribution with mean 300 and standard deviation 25, and then rounded to the nearest integer. \hl{Other parameters of distributional semantics model training were identical to those in \cite{Frank2017}}.  

\subsubsection{Training corpora}
To get representations of English words, the model was trained on the first slice of the ENCOW14 web corpus \cite{Schaefer2015}, comprising 28.9~million sentences with 644.5~million word tokens of 2.8~million types (token and type counts include punctuation, numbers, etc.). This is the same corpus that was used in earlier work to obtain word-vector distances that predict neural activation during sentence reading \cite{Frank2017}.

For Chinese, the model was trained on the Chinese Wikipedia full-text corpus (downloaded 1~November 2016). We used Wikipedia Extractor 
(version 2.66) to extract cleaned text from the downloaded Wikipedia XML. Traditional Chinese text was converted into simplified Chinese by OpenCC (version 0.42).
Chinese is standardly written without explicit word boundaries but these are required by the distributional semantics model. Therefore, the Chinese corpus was segmented into words, using Jieba (version 0.38). Following this, the corpus was converted to pinyin as described under `Materials' above but without manual checking. The resulting  corpus comprised almost 898,000 articles with a total of 210.8~million pinyin word tokens of 2~million types.

\subsubsection{Representing incomplete words}
The Chinese materials include multisyllabic words, for which cortical entrainment was crucially established at the syllabic rate. Capturing this in the model requires a vector representation at each syllable position. However, distributional semantics models generate only vectors that represent complete words. The construction of syllable-level representations is loosely based on the Cohort model of spoken word recognition \cite{MarslenWilson1987}: The syllable sequence $s_1,\ldots,s_n$ (possibly containing only one syllable) activates all words that begin with that sequence; this set of words is called the cohort. A word has to occur at least 5 times in the Wikipedia corpus to be considered part of the cohort. The vector at syllable position $n$, representing the sequence $s_1,\ldots,s_n$, equals the average vector of words in the cohort, weighted by the words' corpus frequencies. The cohort becomes empty when $s_1,\ldots,s_n$ does not form the beginning of any word, in which case syllable $s_n$ starts a new cohort. 
Note that this method can be applied to the sentence sequences as well as the shuffled syllable sequence because it does not depend on (knowledge of) word boundaries. Qualitatively identical results were obtained with a slightly alternative scheme in which cohorts are not included at word-final position, that is, the representation at each word-final syllable equals the single vector for that word. 

\subsection{Lexical information over time}
In  Ding et al.'s Chinese experiments, syllables come in at a fixed rate of 4~Hz, or every 250~ms. The English experiments used a slightly slower presentation rate, but for simplicity we model English and Chinese experiments using the same 4~Hz rate.

Let ${\bf v}=(v_1,\ldots,v_N)$ be the $N$-dimensional column vector that represents the English word or Chinese syllable sequence currently being presented. We assume that the lexical information does not immediately appear at word onset ($t=0$~ms) but some time later, at $\tau\geq 0$. The value of $\tau$ is randomly sampled at each word/syllable presentation, from a uniform distribution with mean $\mu=40$ and width $\beta=50$ (how this choice of parameter values came about is discussed below).

The available lexical information at $t$ milliseconds after word onset (for $0\leq t\leq 250$) is represented by a column vector ${\bf w}(t)=(w_1(t),\ldots,w_{N}(t))$ with
\[
w_i(t) =
\begin{cases}
\varepsilon_i(t) & \text{if } t<\tau\\
v_i + \varepsilon_i(t) & \text{if } t\geq\tau
\end{cases}
\]
where $\varepsilon_i(t)$ denotes Gaussian noise with mean 0 and standard deviation $\sigma=0.5$.
Hence, the lexical information vector ${\bf w}(t)$ starts with representing only noise but at $t=\tau$ the information in ${\bf v}$ becomes available (if $\tau<0$, it becomes becomes available immediately, at $t=0$). 
In practice, we discretize continuous time $t$ into 5~ms bins, corresponding to the 200~Hz low-pass filter frequency applied by Ding et al., so that there are 50 time steps between two syllable onsets. 

All vectors ${\bf w}$ for the stimuli sequence of an experimental condition are concatenated into a single matrix ${\bf W}$ that captures the entire session's time sequence, with a different random order of trials for each of the twelve simulated participants. This matrix has $N$ rows and 50 columns per syllable of the stimulus sequence.

The values of the model's three free parameters ($\mu, \beta$, and $\sigma$) were chosen to obtain results on the English sentences that were visually similar to those of Ding et al., using a different set of word vectors (see Appendix~\ref{parset}).
 
\subsection{Analysis}
Following  Ding et al., we applied a Discrete Fourier Transform to obtain a power spectrum for each experimental condition and each simulated participant. The individual rows of matrix ${\bf W}$, each representing the time course in a single dimension of word vector space, were transformed to the frequency domain. Next, these per-dimension power spectra were averaged over the $N$ dimensions to obtain the power at each frequency bin. Frequency bin width was 1/9~Hz for Chinese and 1/11~Hz for English, as in Ding et al. Again following  Ding et al., we tested whether the power at each frequency bin significantly exceeded the average of the previous and next two bins using one-tailed $t$-tests with false discovery rate correction \cite{Benjamini1995}.

\section{Results}
\begin{figure}
\includegraphics{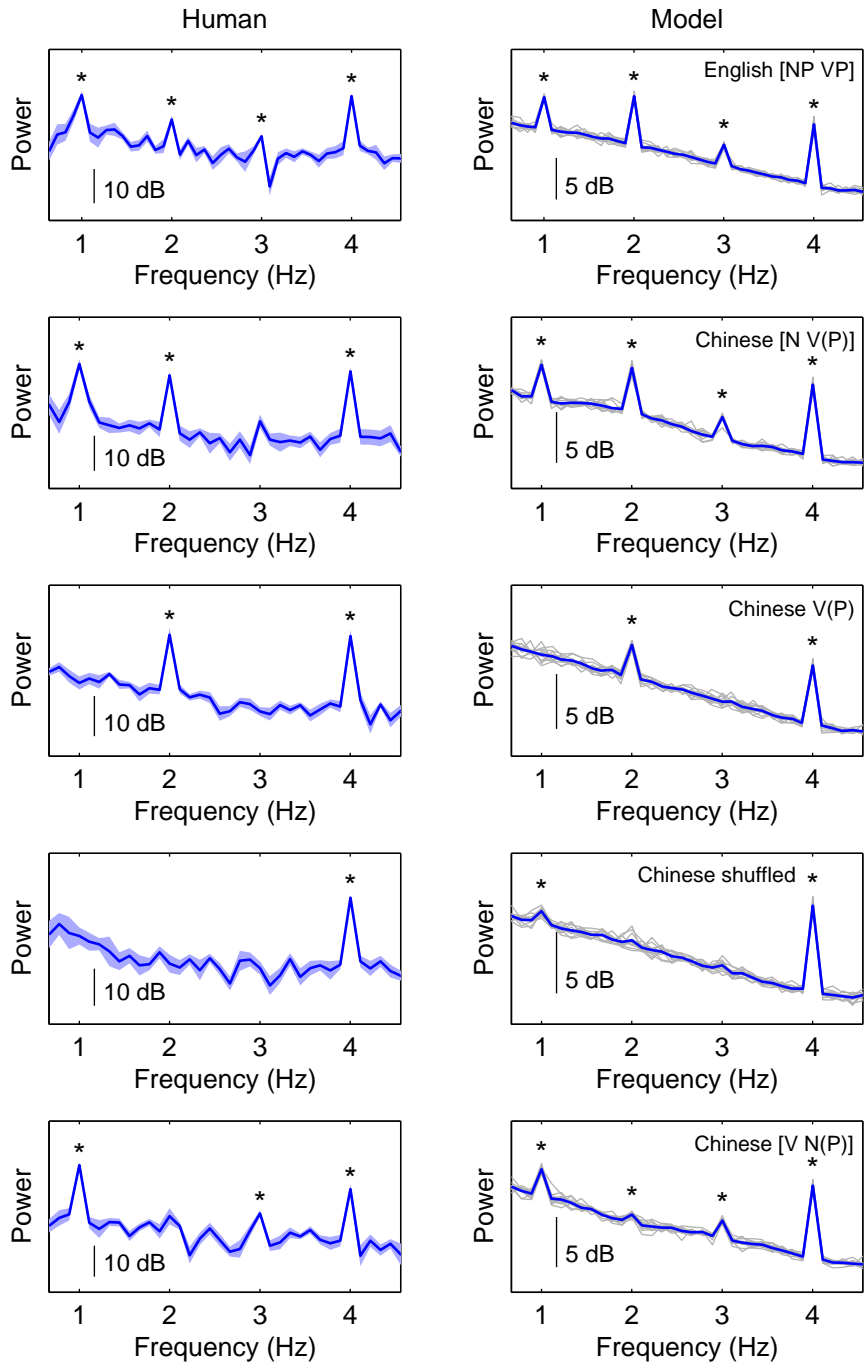}
\caption{Power spectra from human MEG signal (left) and corresponding model predictions (right) in all five conditions. Shaded areas in the human results represent standard errors from the mean over eight subjects. Grey lines in the model results depict individual model runs (simulated subjects). Blue lines are the averages over (simulated) subjects. Statistically significant peaks ($P<.025$; one-tailed) are indicated by asterisks. In the top left panel (human results on English stimuli) the frequency scale has been adjusted to match the simulated presentation rates.} \label{fig:main_results}
\end{figure}

Figure~\ref{fig:main_results} displays the model predictions for all five conditions, side by side with the original results from  Ding et al. Ignoring differences in scale, the two sets of power spectra are strikingly similar. For English four-word [NP VP] sentences, the model predicts peaks in the power spectrum at the presentation rate of words (4~Hz), phrases (2~Hz), and sentences (1~Hz). Results for Chinese four-syllable [N V(P)] sentences look very similar to those for English and, crucially, to those of Ding et al.: Peaks occur at the presentation rates of syllables, words/phrases, and sentences. 

The minor peaks at 3~Hz in both conditions are also visible in the Ding et al. results, although it only reaches significance for the English sentences. Although the 3~Hz peak in the English condition is significant in our reanalysis of Ding et al.'s data, it is not in their original analysis. Possibly, this is because because we applied a less conservative false-discovery rate correction method. The 3~Hz peaks most likely occur as the second harmonic of the 1~Hz signal \cite{Zhou2016} so they do not reflect any interesting property of the input.

The 2~Hz and 4~Hz peaks do not merely arise as harmonics because they remain when only the verbs and verb phrases of the Chinese [N V(P)] sentences are presented, even though the power spectrum lacks the 1 Hz peak in this condition. This is the case both in the model predictions and the neural data. When the syllables are randomly reassigned to sentences, breaking up the word and sentence structure, only the 4~Hz peak should remain. In the model, the 2~Hz and 3~Hz peaks are indeed no longer present and the 1~Hz peak has almost completely vanished: The peak size (defined as peak power minus the average power of the previous and next two frequency bins) is significantly reduced compared to the [N V(P)] sentences (paired $t$-test: $t(11)=26.0; p<.00001$).

Finally, when the stimuli sequence consists of sentences with a one-syllable verb followed by a three-syllable noun or noun phrase, no linguistic unit occurs at a 2~Hz rate. The model results in Figure~\ref{fig:main_results} show that the 2~Hz peak is indeed strongly reduced compared to the [N V(P)] condition (paired $t$-test: $t(11)=30.8; p<.00001$). The corresponding results from  Ding et al. also show a small 2~Hz peak in this condition, although it fails to reach significance. However, note that individual differences between simulated participants are much smaller than between human participants; Hence, very small model effects can reach significance whereas they are washed out by noise in the corresponding human data. Most likely, the small 2~Hz peaks are merely the first harmonic of the 1~Hz signal.

\section{Discussion}
For all five stimuli types, our model predicts power spectra that are qualitatively almost identical to those from Ding et al.'s MEG study. 
Only the very small but statistically significant 1~Hz peak in the Chinese shuffled syllable sequence condition was not present in the human results. Apparently, some aspects of the syllables still recur at a 1~Hz rate, albeit very weakly. This is in fact not surprising considering that the syllables kept their original position in the sentence. A possible explanation for the difference between model prediction and human data is that the model treats the shuffled syllable sequence and grammatical sentences exactly the same, in that  at each point all possible word candidates are selected (see under `Representing incomplete words' in the Methods section). In contrast, human participants are likely to forgo pro-active word activation when listening to non-word sequences, even though their task in this condition was to detect the occasional correct sentence. Crucially, however, the predicted 1~Hz peak was much reduced compared to the [N V(P)] sentence condition, which shows that shuffling the syllables affects the model predictions in the same direction as the MEG power spectra. 

\subsection{Lexical versus structural accounts of the Ding et al. results}
The model's only linguistic representations are formed by the distributional semantics vectors.
Although these are learned from word strings (see Methods), they do not explicitly encode information about word sequences (such as transitional probabilities) and can therefore not be used to predict, for example, that a noun is likely to be followed by a verb \cite{Frank2017}.
More importantly, a word's vector does not depend on its position or neighbors in the stimulus sentence, and the model lacks higher-level representations of phrases and sentences. 
This, of course, raises a crucial question: What is the origin of the predicted power peaks at the presentation rates of phrases and sentences?

The word vectors represent lexical properties by virtue of the fact that similarities between vectors mirror paradigmatic relations between words: Words that share more syntactic/semantic properties are encoded by more similar vectors. Consequently, if certain lexical properties occur at a fixed rate in the stimulus sequence, this will be reflected as a recurring approximate numerical pattern in the model's time-series of vectors. For example, in  Ding et al.'s English four-word sentences condition, every other word is a noun, most often referring to some entity, and every forth word is a transitive verb, usually referring to an action. Because semantically and syntactically similar words are represented by similar vectors, the vector sequence corresponding to this experimental condition shows spectral power peaks at exactly the occurrence rates of two-word phrases and four-word sentences. Crucially, this does not rely on any hierarchical structure or process: Vectors represent only lexical information and the spectral power analysis is applied over a sequence of vectors that is not processed, integrated, or interpreted. Note, however, that whether a word is a noun or verb (or, in terms of semantics: refers to an entity or action) depends on its role within the sentence. In the stimulus sentence ``fat rat sensed fear'', for example, the individual word ``fat'' could be a noun instead of an adjective and ``fear'', on its own, could be a verb instead of a noun. The vectors do not distinguish between the different senses of these ambiguous words: There is only one representation of ``fat''. Nevertheless, there is apparently enough repetition in the stimulus word's properties to account for the MEG results. 
\hl{In fact, we obtained qualitatively similar results when each word was represented by a vector that merely identifies its most frequent syntactic category, that is, independently of the word's role in the sentence (see Appendix~\ref{vecs_pos}).}

\begin{figure}
\includegraphics{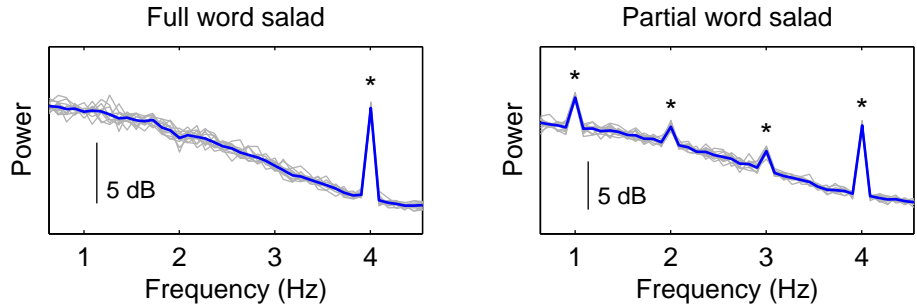}
\caption{Model predictions on word salad conditions. Stimuli sequences were constructed by randomly drawing (with replacement) words from the English [NP VP] sentences. Left: all words randomly drawn. Right: adjectives and verbs keep their original position in the [NP VP] stimuli.}
\label{fig:wordsalad_results}
\end{figure}

The model shows how Ding et al.'s cortical entrainment results can be explained without recourse to hierarchical structures or integrative processes. However, this does not rule out that cortical entrainment to hierarchical structure exists or even that the  Ding et al. results in fact do reflect processing of hierarchical syntax. Indeed, it has recently been demonstrated that a hierarchical sentence processing model predicts similar power spectra, at least on English materials \cite{Martin2017}. 
Whether lexical or phrasal/sentential properties of the speech signal are in fact responsible for cortical entrainment can possibly be established by testing on a `word salad' version of the English stimuli sequence, created by keeping all verbs and adjectives in place but replacing all nouns by random words from different syntactic categories. This breaks any (consistent) syntactic structure so no 1~Hz peak should be visible if its occurrence depends on the repeated presence of 4-word sentences. In contrast, our lexical model does predict 1~Hz power in this word-salad condition because both verbs and adjectives still occur at a 1~Hz rate (see Figure~\ref{fig:wordsalad_results}).

Returning to the stimuli sequences tested by Ding et al., our model demonstrates that hierarchical syntax is not necessary to explain these entrainment results: Representing only lexical properties of the stimuli suffices.
Compared to structural accounts (such as \cite{Martin2017}), our lexical explanation has the advantage of parsimony. This is because 
constructing a sentence's hierarchical structure requires information about the words' \hl{possible} syntactic categories (i.e., whether a word \hl{can be} a noun, verb, or adjective)\hl{; or in semantic terms: Understanding a sentence requires at least some knowledge of word meaning.} Thus, lexical properties are necessary in a structural explanation of the entrainment results. Conversely, however, \hl{the vector representations in our model do not depend on the sentence in which the words happen to appear. Hence, the lexical model} forms a more parsimonious account.

\subsection{Word vectors as cortical representations}
The vector representations are not intended to be neurally realistic, that is, we do not claim they are isomorphic to cortical representations. Rather, the relation between model and brain is a higher-order one: The rhythmic recurrence of patterns in a vector sequence corresponds to the patterns in the MEG signal caused by perceiving the stimuli represented by the vectors. Ding et al. report two additional experiments that probe the cortical representations more directly. First, they show that overall MEG activity strongly decreases around phrase and sentence boundaries. Second, they find a spatial dissociation between the cortical areas that show a phrasal- or sentential-rate response.  Although there was no reason to expect our vectors to account for these findings too, we did find that vector lengths show a sharp drop right after phrase and sentence boundaries (Appendix~\ref{vectorlength}) and that (apparent) effects of phrases and sentences can be localized in different vector dimensions (Appendix~\ref{localization}). Hence, the correspondence between the distributional semantic vectors and cortical representations may be stronger than anticipated.

\subsection*{Acknowledgments}
We are very grateful to Nai Ding for sharing his data and to Jona Sassenhagen, Tal Linzen, and two anonymous reviewers for their comments on earlier versions of this paper. The research presented here was funded by the Netherlands Organisation for Scientific Research (NWO) Gravitation Grant 024.001.006 to the Language in Interaction Consortium.


\pagebreak
\appendix

\section{Representing words as syntactic categories} \label{vecs_pos}
\hl{Under the model's explanation of the Ding et al.'s results, similar outcomes should be obtained if vectors represent only the syntactic categories of the stimulus words. The question remains if this can be done without interpreting the stimuli beyond the individual words, because for many words the syntactic category (also known as part-of-speech; POS) is ambiguous until the word is understood in its sentence context.}

\begin{figure}
\includegraphics{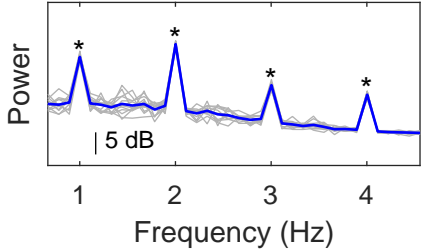}
\caption{Power spectra resulting from processing English [NP VP] sentences with each word replaced by its most frequent syntactic category.} \label{fig:vecs_pos}
\end{figure}

As a first test of the feasibility of a `POS only'-account, we replaced each word of the English [NP VP] stimuli by its most frequent POS in the ENCOW corpus (i.e., without considering the word's role in the sentence), yielding 13 different POS tags. Each was assigned a vector such that all 13 vectors are orthogonal, that is, they identify POS without encoding any notion of similarity between syntactic categories. As much as possible without sacrificing orthogonality, vector values were randomly drawn from the original vectors (i.e., representing words), again for 12 simulated participants. The resulting frequency spectra (Figure~\ref{fig:vecs_pos}) again shows the power peaks at 1, 2, and 4~Hz. This suggests that it is possible that POS representations underlie the MEG findings. However, note that this result should not be taken as evidence for the cognitive or neural representation of syntactic categories (let alone for the particular set of POS tags in the ENCOW corpus) because similar outcomes were obtained using the original vector representations in which words only approximately cluster by POS.

\section{Vector lengths over the course of a sentence} \label{vectorlength}
Ding et al. presented participants with 90 Chinese sentences consisting of a 3- or 4-syllable NP followed by a 4- or 5-syllable VP. Overall MEG activity decreased strongly near the NP and sentence boundaries, which was taken as further evidence for neural tracking of phrasal and sentential structure. 

\begin{figure}
\includegraphics{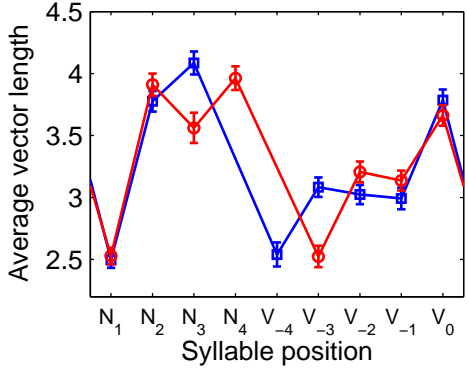}
\caption{Average euclidean length of vector representations at each point of 7- and 8-syllable Chinese [NP VP] sentences. Error bars indicate 95\% confidence intervals. Plots are left aligned on the 3- or 4-syllable NP (labeled N$_1$ to N$_4$) as well as right aligned on the 4- or 5-syllable VP (labeled V$_{-4}$ to V$_0$). Sentence with a 3-syllable NP have no N$_4$ and 4-syllable NP sentences have no V$_{-4}$.} \label{fig:N3_N4_sim}
\end{figure}

We obtained vector representations for pinyin versions of the same stimuli and took each vector's euclidean length as a measure for overall `activation'. Figure~\ref{fig:N3_N4_sim} plots the average length as a function of syllable position in the sentence. There are indeed sharp drops after NP and sentence (VP) boundaries, corresponding to the human results. This is most likely caused by the fact that syllable strings that cross a phrase boundary rarely form a possible word beginning. Consequently, the first syllable of a phrase often starts a new cohort, and averaging over its many words yields a relatively short vector.

\section{Localization of sentential- and phrasal-rate responses} \label{localization}
In an electrocorticography (ECoG) study using the English four-word [NP VP] sentences, Ding et al. identified the electrodes that selectively respond at the phrasal or sentential rate but not at the word-presentation rate. Across these electrodes, power at the phrasal and sentential frequencies were negatively correlated, which ``demonstrates spatially dissociable neural tracking of the sentential and phrasal structures'' (p. 162).

For each individual dimension of vector representations for the same stimuli sequences, we computed the peak size at each frequency as the power at that frequency minus the average over all frequencies within 0.5~Hz on either side. These peak size were then expressed as $z$-scores relative to the range from 0.5 to 4.5~Hz. Next, we identified vector dimensions that show a sentential- or phrasal-rate pattern but no word-rate pattern by selecting those with $z>2$ at 1 or 2~Hz and $z<1$ at 4~Hz. Across the 12 simulated subjects, 10.0\% of vector dimensions met these criteria. 
On these dimensions, peak sizes at 1 and 2~Hz were negatively correlated ($r=-.43; p<.00001$) which shows that the lexical properties represented by the vectors are dissociated across vector dimensions in a way that corresponds to the the spatial dissociation between sentential- and phrasal-rate responses in Ding et al.'s ECoG experiment.

\section{Setting model parameters} \label{parset}
The model has three free parameters: 
\begin{itemize}
\item $\mu$: The average arrival time of lexical information, in simulated milliseconds from word onset
\item $\beta$: The width of the uniform distribution of arrival time of lexical information
\item $\sigma$: The standard deviation of Gaussian noise
\end{itemize}
Appropriate values for these parameters were found by getting a new set of word vectors (with $N=300$) for the English [NP VP] sentences. The objective was to find $\mu, \beta$, and $\sigma$ such that the frequency spectrum resembles the corresponding results from Ding et al. in this condition, that is, there are peaks at 1~Hz, 2~Hz, and 4~Hz, possibly a minor peak at 3~Hz, no other peaks, and a slight increase in power for lower frequencies.

\begin{figure}
\includegraphics{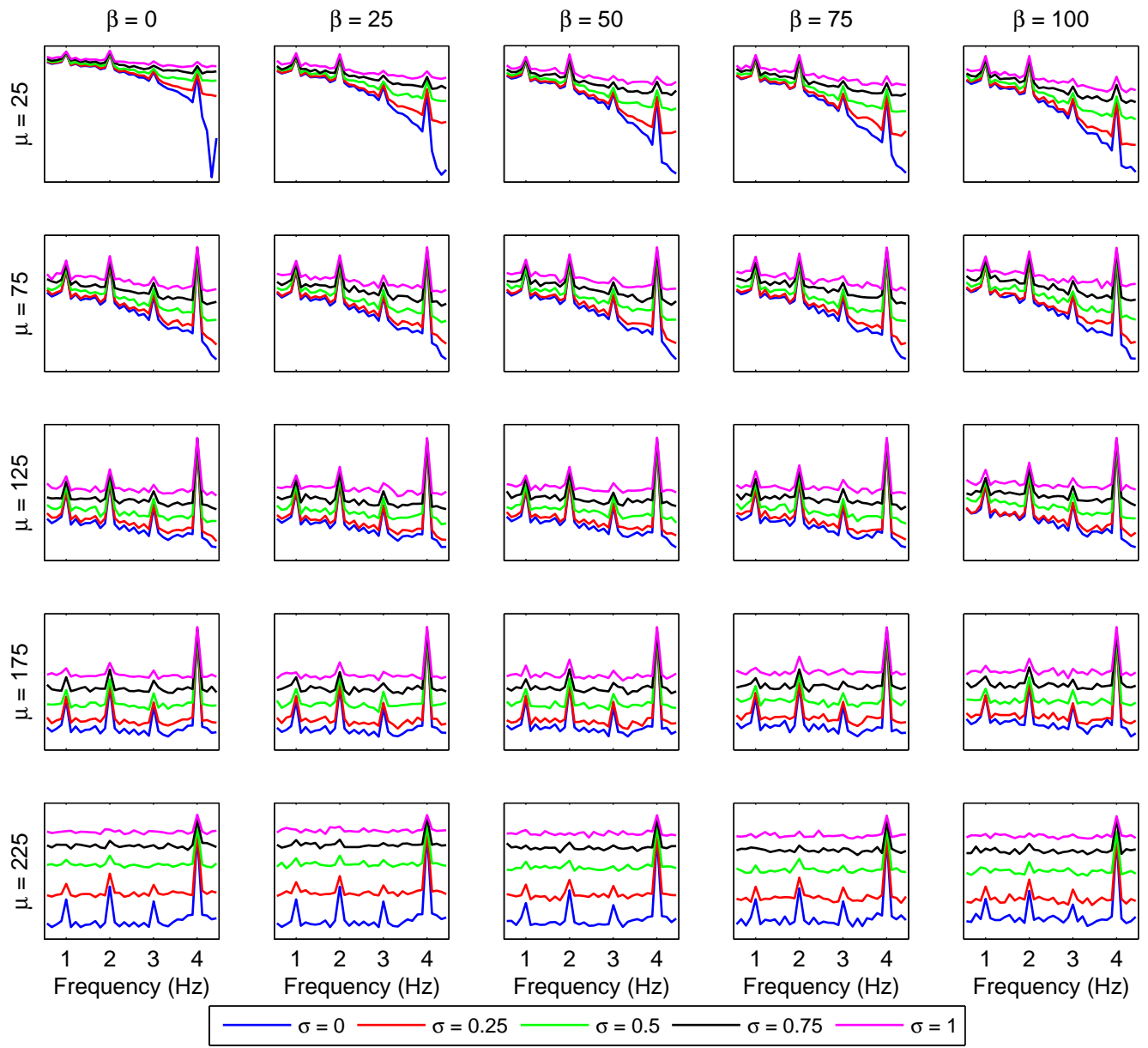}
\caption{Power spectra resulting from processing English [NP VP] sentences, for different combinations of parameter values.} \label{fig:parset1}
\end{figure}

\begin{figure}
\includegraphics{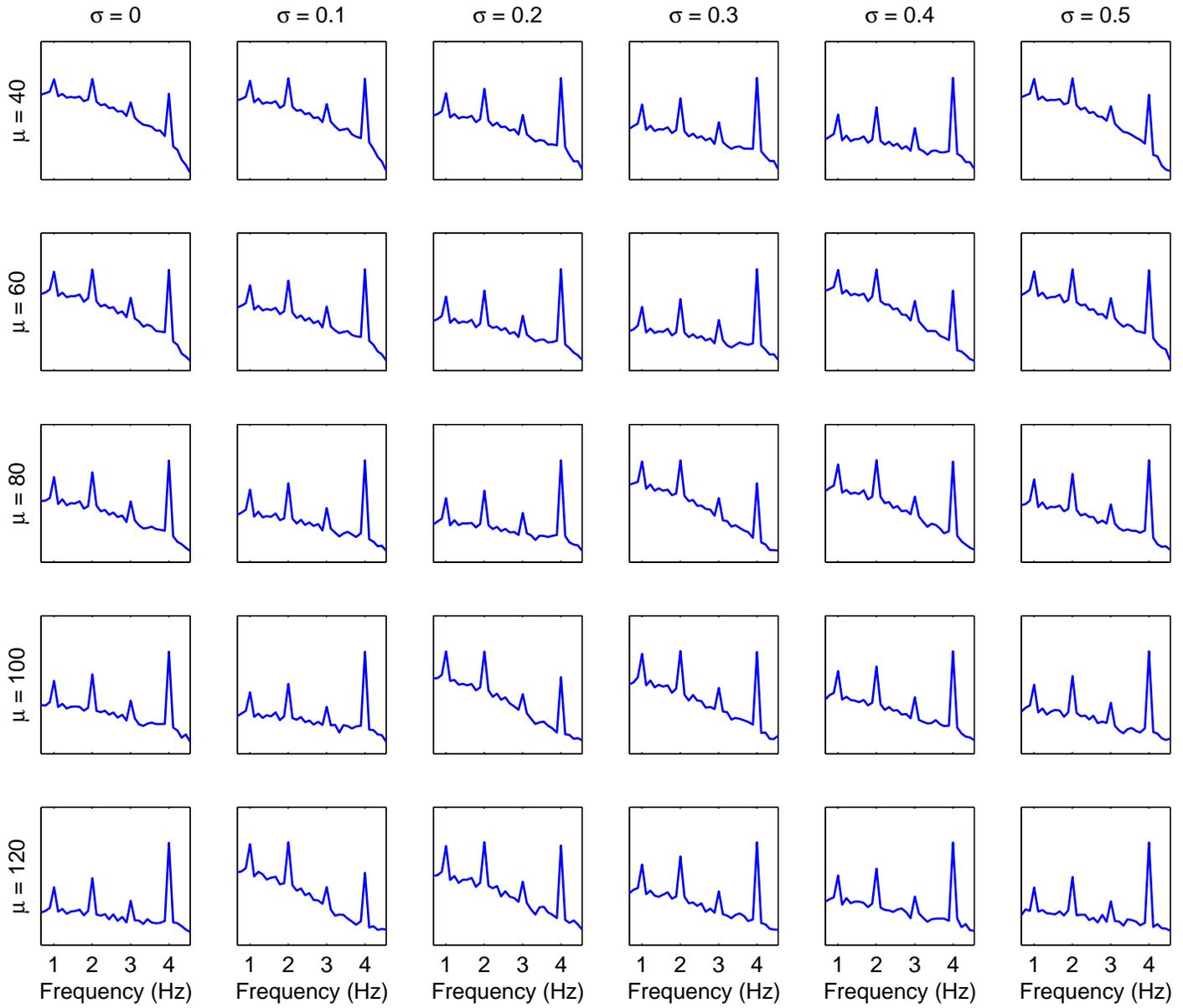}
\caption{Power spectra resulting from processing English [NP VP] sentences, for $\beta=50$ and different values of $\mu$ and $\sigma$.} \label{fig:parset2}
\end{figure}

Figure~\ref{fig:parset1} shows the power spectra for each combination of $\mu\in\{25,75,125,175,225\}, \beta\in\{0,25,50,75,100\}$, and $\sigma\in\{0,.25,.5,.75,1\}$. The three major peaks at 1~Hz, 2~Hz, and 4~Hz, and the minor peak at 3~Hz, are clearly visible for most combinations of parameter values. Only for very high levels of noise $\sigma$ or when both $\mu$ and $\beta$ are high (which means that much lexical information never arrives) are the power spectra mostly flat. These results motivated us to fix $\beta$ at 50 and perform a more fine-grained search through parameter space for low-to-medium levels of $\mu$ and $\sigma$. Based on visual inspection of the resulting power spectra in Figure~\ref{fig:parset2}, we settled for $\mu=40$ and $\sigma=0.5$. The parameters are not over-fitted, as is clear from the fact that the same combination of values resulted in very similar power spectra using another set of vectors as well as on the Chinese sentences (Figure~\ref{fig:main_results}).

\end{document}